\documentclass[aps,pra,reprint,groupedaddress]{revtex4-1}

\input{epsf}

\usepackage[pdftex]{graphicx}
\usepackage{amsmath} 

\begin{document}




\author{G.~G.~Kozlov}
\affiliation{Spin-Optics laboratory, St.~Petersburg State University, 198504 St.~Petersburg} 

\author{I.~I.~Ryzhov}
\affiliation{Spin-Optics laboratory, St.~Petersburg State University, 198504 St.~Petersburg}

\author{V.~S.~Zapasskii}
\affiliation{Spin-Optics laboratory, St.~Petersburg State University, 198504 St.~Petersburg}







\begin{abstract}

A strict analytical solution of the problem of spin-noise signal formation in a volume medium with randomly moving spin carriers is presented. The treatment is performed in the model of light scattering in a medium with fluctuating inhomogeneity.   Along with conventional single-beam, geometry, we consider the two-beam arrangement, with the scattering field of the auxiliary ("tilted") beam heterodyned on the photodetector illuminated by the main beam. It is shown that the spin noise signal detected in the two-beam arrangement is highly sensitive to motion (diffusion) of the spin carriers within the illuminated volume and thus can provide additional information about spin dynamics and spatial correlations of spin polarization in volume media. Our quantitative estimates show that, under real experimental conditions, spin diffusion may strongly suppress the spin-noise signal in the two-beam geometry. Mechanism of this suppression is  similar to that of the time-of-flight broadening with the critical distance determined by the period of spatial interference of the two beams rather than by the beam diameter.

\vskip10mm
 

\vskip10mm

\end{abstract}

\title{Spin noise spectroscopy of randomly moving spins in the model of light scattering:
 Two-beam arrangement}
\maketitle

\date{\today}


\section*{Introduction }

Spectroscopy of spin noise rapidly developing during the last decade has shown itself as an efficient method of 
research with a wide range of interesting informative abilities in the field of 
magneto-spin physics \cite{Zap1,Oest2,Sin}.
 The spin noise spectroscopy (SNS)  made it possible to study resonance magnetic susceptibility of nano-objects 
(quantum wells, quantum dots), hardly accessible for the ESR technique \cite{Glazov,singlehole}, 
to observe dynamics of nuclear magnetization \cite{R,R1}, and to investigate certain nonlinear phenomena 
in such systems \cite{R2}. The fact that magnetization is detected, in the SNS, by optical means \cite{f1}, 
provides this method with additional informative channels.  Specifically, studying the spin-noise power 
dependence on the probe light wavelength makes it possible to identify the type of broadening 
(homogeneous/inhomogeneous) of optical transitions \cite{Zap2,TC}. Temporal modulation of the probe beam 
(e.g., shaping the ultrashort optical pulses) allows one to extend the range of the detected noise signals up 
to microwave frequencies \cite{To1}. The use of tightly focused probe beams provides oportunity  of detecting the noise 
signals with a high spatial resolution and even to perform 3D-tomography of magnetic properties of materials \cite{To}.
 The range of objects of the SNS is not restricted to solid-state systems. Nowadays, this method is widely applied 
to studying atomic gases \cite{Mitsui} from which the history of the SNS has been started \cite{Zap}.

Magnetic state of a material (magnetization), in the SNS, is monitored by polarization plane rotation of the probe 
beam transmitted through the sample. It is assumed, in these measurements, that the detected angle of the polarization
 plane rotation is proportional to total magnetization of the illuminated volume of the sample. This is considered 
to be valid even for spontaneous spatiotemporal stochastic fluctuations of the magnetization detected in the SNS 
This simple picture is commonly used to interpret experimental data on SNS. In a consistent analysis, however, 
polarimetric signal detected in the SNS should be regarded as a result of scattering of the probe light by the 
randomly gyrotropic medium \cite{Gorb}. Such an analysis performed in \cite{Koz} allowed us to justify 
the above simple picture 
and, besides, to propose a two-beam modification of the SNS that makes it possible to observe both temporal and 
spatial correlations in magnetization of the illuminated region of the medium. In \cite{Koz}, we restricted our 
treatment 
to the case of thin samples (compared with the Rayleigh length of the probe beam), typical for experiments with 
solid-state samples. In this paper, we consider a more general case of a volume media with moving spin carriers 
(more typical for atomic vapors).  In the first part of the paper, which is a continuation of publication \cite{Koz}, 
we analyze formation of the SNS signals for the samples with the thickness exceeding Rayleigh length of the 
focused light beams.  We show that the noise signal ceases to increase with the sample thickness when it 
substantially exceeds the Rayleigh length. For the case of two-beam arrangement, we derive an explicit 
expression for the spin-noise signal in the medium with spin diffusion. Our estimates show that atomic 
diffusion in gaseous systems may drastically suppress the noise signal created by the auxiliary beam and, 
thus, hinder its observation.

The paper is organized as follows. In Sect. 1, in the single-scattering approximation, we derive the expression for the noise polarimetric signal from the sample transilluminated by two coherent laser beams (referred to as {\it main} and {\it auxiliary}), with only one of them (the main) hitting the detector (Eq. (\ref{13})).  In Sect. 2, we obtain relationships for the gyrotropy noise power spectrum detected in the SNS. We present calculations of these spectra for the samples of arbitrary thickness in the framework of the model of resting gyrotropic particles and of the diffusion model. We show that amplitude of the noise spectrum is getting independent of the sample thickness when the latter exceeds the Rayleigh length of the beam (Eq. (\ref{20})).  In this Section we also describe the effect of ‘time-of-flight’ broadening  of the spectrum arising in the diffusion model  and present a simple experimental illustration of the made conclusions using as a model object  a thick cell with Cs atoms in a buffer-gas atmosphere. In Sect. 3, we present analysis of signals observed in the two-beam arrangement of SNS \cite{Koz}.  For the contribution to the noise spectrum associated with the auxiliary beam, we obtain expression that takes into account diffusion of the gyrotropic particles (Eq. (\ref{37})).   
Recommendations are given regarding  the choice of the systems where the above signal can be observed. The results of the work are summarized in Conclusions.

\section{ Polarimetric signal from a randomly gyrotropic sample: the two-beam arrangement}

\begin{figure}
 \centering
\includegraphics[width=\columnwidth,clip]{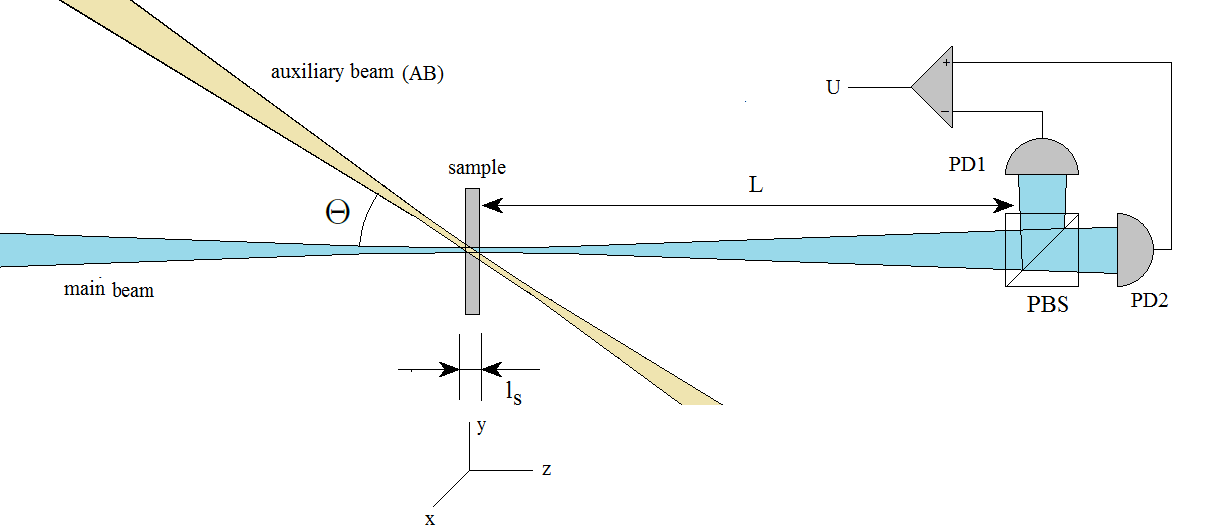}
 \caption{The two-beam experimental arrangement. PBS -- polarization beamsplitter, PD1 and PD2 -- photodetectors. }
 \label{fig1}
\end{figure} 

In this section, we present solution of a problem typical for the noise spectroscopy. 
Let us consider a weakly gyrotropic sample with  spatial distribution of the gyration vector 
described by the function ${\bf G(R)}$, with $|{\bf G(R)}|\ll 1$. The sample is probed with two Gaussian beams with a 
frequency $\omega$ (Fig. 1), for which the sample is transparent.  One of the beams (further referred 
to as {\it main}), after passing through the sample, hits the differential polarimetric detector 
comprised of a polarization beamsplitter (PBS) and two photodetectors PD1 and PD2.  The total output 
signal is obtained as a difference of  signals of the detectors PD1 and PD2. The second beam 
(further referred to as {\it auxiliary}) also passes through the sample, but does not hit the detector.
 Electric fields of the main and auxiliary beams will be denoted, respectively, as 
${\bf E_0 (R)}$ and ${\bf E_0^t(R)}$. We 
assume that the detector is initially balanced , i.e. polarization of the main beam is chose so that, 
in the absence of the sample (at ${\bf G(R)}\equiv 0$), the output signal of the detector is zero. Our task is to 
find the gyrotropy-related increment of the output signal  $\delta U$ (in what follows - just  {\it signal}) 
in the first order of gyrotropy ${\bf G(R)}$. A similar problem for thin (compared with the Rayleigh length) samples 
was considered in \cite{Koz}.  Below, we present solution of this problem for samples of arbitrary length.

The signal $\delta U$ arises due to the fact that at ${\bf G(R)}\ne 0$ the beam hitting the detector 
 contains not only the 
field of the main beam, but also the field ${\bf E_1(R)}$ that appears as a result of scattering of the main or 
auxiliary beam by the sample. Since we neglect any optical nonlinearity,  these two fields may be 
calculated independently, and the signal $\delta U$ may be represented as a sum of two contributions 
related to scattering of the main and auxiliary beams. Since the detector is permanently irradiated by 
the main beam, detection of these fields occurs in the regime of heterodyning, with the role of local 
oscillator played by the field of the main beam.

In what follows, we will use complex electromagnetic fields with time dependence in the form $ e^{-\imath\omega t}$ 
assigning physical sense to their real parts (which will be denoted by calligraphic letters).  
The calculations will be performed in the coordinate system with its $x$- and $y$-axes aligned along 
principal  directions of the polarization beamsplitter and $z$-axis collinear with the main beam. 
The coordinate  origin is located in the region of the sample, with its characteristic size 
$l_s$ being much smaller than the distance from the photodetector $L$:  $l_s\ll L$ (Fig. 1). For the 
signal $\delta U$, we are interested in, we will use the following expression \cite{Koz}:
\begin{widetext}
        \begin{equation}
 \delta U={\omega\over\pi}\hbox { Re }\int _{0}^{2\pi/\omega} dt\int_{-l_x}^{l_x}dx\int_{-l_y}^{l_y}dy 
\bigg [
{\cal E}_{x0}(x,y,L) E_{x1}(x,y,L)-{\cal E}_{y0}(x,y,L) E_{y1}(x,y,L)\bigg ]
\label {1} 
\end{equation} 
\end{widetext}
Here, the integration over $x$ and $y$, for products of components of the complex field of scattering ${\bf E_1}(x,y,L)$ 
and real part of the field of the main beam   ${\bf {\cal E}}(x,y,L)\equiv$ Re $ {\bf E_0} (x,y,L)$, is performed over the effective photosensitive 
surface of the detector $2l_x\times 2l_y$ located at a distance $L$ from the sample along the main beam 
propagation direction (Fig. 1). The integration over $t$ corresponds to averaging over the period 
of optical oscillations.

Below, we, following \cite{Koz}, will calculate the field of scattering produced by the auxiliary beam 
(we will denote it, as before, by ${\bf E_1(R)}$) and the related polarimetric signal denoted by $\delta U_t$. 
As shown in \cite{Koz}, this field satisfies the inhomogeneous Helmholtz equation                                                               

\begin{equation}
  \Delta {\bf E_1}+k^2{\bf E_1}=-4\pi k^2\alpha({\bf r}){\bf E_0^t(r)}\equiv -4\pi k^2 {\bf P^t(r)} .
  \label{2}
  \end{equation}

    Here, $k\equiv \omega/c$ ($c$ is the speed of light),  $\alpha ({\bf r})$ is the polarizability 
tensor of the gyrotropic 
medium (connected with the gyration vector as $\alpha_{ik}({\bf r})=\imath\varepsilon_{ikj}G_j({\bf r})$ where
$\varepsilon_{ijk}$-- unit antisymmetric tensor), 
and ${\bf P^t(r)}$ is the sample polarization induced by the 
field   ${\bf E_0^t(R)}$ of the auxiliary beam.

Solution of Eq. (\ref{2}) is obtained using the Green function  $ \Gamma({\bf r}) $ of the Helmholtz 
operator $ \Gamma({\bf r})=-{e^{\imath kr}/4\pi r}$ and has the following form  

    \begin{equation}
       {\bf E_1(r)}=k^2\int  {e^ {\imath k|{\bf r-R}|}\over {\bf |r-R|}}  {\bf P^t(R)}
       d^3{\bf R}
       \label{3}
        \end{equation}

For further calculations, it is convenient to introduce the vector function ${\bf \Phi( R})$ with the components 
defined by the expression  \cite{f0}
\begin{equation}
  \Phi_i({\bf R})\equiv \int_S dxdy\hskip2mm 
          {\cal E}_{i0}(x,y,z)
           {e^ {\imath k|{\bf r-R}|}\over {\bf |r-R|}}\bigg |_{z=L} 
\label{4}
  \end{equation}
where $ {\bf r}=(x,y,z), i=x,y$
 and auxiliary functions $\Phi_{i}^\pm({\bf R})$:

 \begin{equation}
 \Phi_i({\bf R})\equiv \Phi_{i}^+({\bf R})e^{-\imath\omega t}+\Phi_{i}^-({\bf R})e^{\imath\omega t}
 \label{5}
 \end{equation}

Using Eqs. (\ref{1}), (\ref{3})  and (\ref{4}), we can obtain the following equation for the contribution $\delta U_t$
 into the output signal associated with the auxiliary beam

  \begin{equation}
  \delta U_t=k^2{\omega\over\pi}\hbox { Re }\int _{0}^{2\pi/\omega} dt\int d^3{\bf R}
 \bigg [P^t_x({\bf R})\Phi_x({\bf R})-
 \label {6} 
 \end{equation} 
$$
-P^t_y({\bf R})\Phi_y({\bf R})\bigg ]
$$
By substituting ${\bf \Phi(R)}$ into this equation in the form of Eq. (\ref{5}) 
and taking into account that ${\bf P^t(R)}\sim e^{-\imath\omega t}$, we can ensure that, after integration 
over time, only terms containing $\Phi^-_{x,y}({\bf R})$  survive in Eq. (\ref{6}):

  \begin{equation}
  \delta U_t=2k^2\hbox{ Re }\int d^3{\bf R}\bigg [\Phi_x^-({\bf R})P_x^t({\bf R})-\Phi_y^-({\bf R})P_y^t({\bf R})\bigg ]\hskip1mm e^{\imath\omega t}
  \label{7}
  \end{equation}

The factor $e^{\imath\omega t}$ eliminates time dependence of the field ${\bf P^t(R)}$. 
Let us write out explicit expressions for the fields of the main ${\bf E_0(r)}$ and auxiliary ${\bf E_0^t(r)}$ 
Gaussian beams \cite{Koz}

\begin{widetext}
  \begin{equation}
                      {\bf E_0(r)}=e^{\imath (kz-\omega t)}  
                      \sqrt{8 W\over c}
                      {kQ\over (2k+\imath Q^2z)}
                      \exp\bigg [-{kQ^2(x^2+y^2)\over 2(2k+\imath Q^2z)}\bigg ]
                      {\bf d}\equiv {\bf A_0(r)} e^{-\imath\omega t},
                      \label{8}
                      \end{equation}

   \begin{equation}
                   {\bf E_0^t(r)}=e^{\imath (kZ-\omega t+\phi_t)}  
                   \sqrt{8 W_t\over c}
                   {kQ\over (2k+\imath Q^2Z)}
                   \exp\bigg [-{kQ^2(X^2+Y^2)\over 2(2k+\imath Q^2Z)}\bigg ]
                   {\bf d_t} \equiv {\bf A_0 ^t(r)} e^{-\imath\omega t}
                   \label{9}
                   \end{equation}
 \end{widetext}
 where
                               \begin{equation}
                               \left(\begin{matrix} X\cr Y\cr Z \end{matrix}\right )\equiv\hat R {\bf r}+{\bf \delta r},  \hskip5mm 
                               \hat R\equiv \left(\begin{matrix} 1 & 0& 0 \cr 0 & \cos\Theta &\sin\Theta \cr 0 & -\sin\Theta & \cos\Theta \end{matrix}\right )
                               \label{10}
                               \end{equation}

Here,  $W$ and $W_t$ are intensities of the main and auxiliary beams, respectively. The parameter $Q$ 
is connected with the beam radius in the waist $\rho_c$  by the relation $Q\equiv 2/\rho_c$. 
Polarization of the main and auxiliary beams is specified by the Jones vectors ${\bf d}$ and ${\bf d_t}$ 
lying in the planes perpendicular to propagation directions of the beams. The sense of the angle $\Theta$ 
is made clear by Fig. \ref{fig1}, and, as in \cite{Koz}, we assume that $\Theta < 1$.  
The parameters $\delta {\bf r}$ and $\phi_t$   describe, respectively, the spatial and phase shifts of the 
auxiliary beam with respect to the main one. In Eqs. (\ref{8}),(\ref{9}), we introduced time-independent 
amplitudes of the fields of the main and auxiliary beams ${\bf A_0(r)}$ and  ${\bf A_0^t(r)}$. 
Using Eq. (\ref{2}) to express polarization ${\bf P^t(R)}$ through the field of the auxiliary beam (\ref {9}), 
we obtain, with the aid of (\ref{7}), the expression for the detected signal:
\begin{widetext}
\begin{equation}
  \delta U_t
  =2k^2\hbox{ Re }\int d^3{\bf R}\bigg [\Phi_x^-({\bf R})\alpha_{xx}({\bf R})A_{0x}^t({\bf R})+
   \Phi_x^-({\bf R})\alpha_{xy}({\bf R})A_{0y}^t({\bf R})-
\label{11}
  \end{equation}
  $$
  -\Phi_y^-({\bf R})\alpha_{yx}({\bf R})A_{0x}^t({\bf R})
   -\Phi_y^-({\bf R})\alpha_{yy}({\bf R})A_{0y}^t({\bf R})
   \bigg ]
   $$
\end{widetext}
Now, we use the result of \cite{Koz1} showing that the function  $\Phi_i^-({\bf R})$ can be expressed 
through the main beam amplitude ${\bf A_0(R)}$ as follows (see remark \cite{f0})

     \begin{equation}
     \Phi^-_i({\bf R})=-{\imath\pi\over k}A^\ast_{0i}({\bf R})\hskip10mm i=x,y
   \label{12}
     \end{equation}

By substituting Eq. (\ref{12})   into Eq. (\ref{11}) and taking into account that, in the considered case of gyrotropic sample, 
the polarizability tensor has the form $\alpha_{ij}=\imath\varepsilon_{ijk}G_k({\bf R})$ 
($\varepsilon_{ijk}$ is the unit antisymmetric tensor), we obtain the following final expression for the 
polarimetric signal $\delta U_t$ from the gyrotropic sample illuminated by the main and auxiliary light beams:
\begin{widetext}
      \begin{equation}
      \delta U_t=2\pi k\hbox{ Re }\hskip1mm  \int d^3{\bf R}\bigg [
         A_{0x}^\ast({\bf R})A_{0y}^t({\bf R})+A_{0y}^\ast ({\bf R})A_{0x}^t({\bf R})    \bigg ]\hskip1mm G_z({\bf R})
          \label{13}
      \end{equation}
      \end{widetext}
    
     Equation (\ref{13}) shows that the polarimetric signal associated with the auxiliary beam (${\bf A_0^t(R)}$), 
detected by its mixing with the  wave of the main beam (${\bf A_0(R)}$) 
is controlled by gyrotropy of the sample only {\it in the region of overlap} of the two beams. 
Remind that Eq.(\ref{13}) describes contribution to the polarimetric signal arising due to scattering 
of the auxiliary beam.  Along with this contribution, there always exists the contribution related to 
scattering of the main beam observed in the conventional single-beam arrangement, when the auxiliary beam is absent. 
To calculate this contribution, one has just to set 
 ${\bf A_0^t(R)}={\bf A_0(R)}$ in Eq.  (\ref{13}). 
The total signal in the two-beam arrangement is obtained by summation of the two contributions.

\section{The noise power spectrum in the single-beam arrangement}

In this section, we calculate the spin-noise signal for the conventional single-beam geometry.  
Polarization of the main beam (which is the only one in this arrangement) is specified by the Jones 
vector ${\bf d}=(\cos\phi,\sin\phi,0)$  (in the coordinate system introduced above). Using Eq. (\ref{8}), we can 
show that dependence of the beam radius $\rho(z)$ (at e-level of the field squared) on the coordinate $z$ has the form

\begin{equation}
   \rho(z)\equiv \sqrt{4k^2+Q^4z^2\over 2k^2Q^2}={\rho_c\over \sqrt 2}\sqrt{1+{z^2\over z_c^2}}
  ={\lambda \over \sqrt 2 \pi \rho_c}\sqrt{z_c^2+z^2}
  \label{14}
   \end{equation}

Here, $\rho_c=2/Q$, $\lambda\equiv 2\pi/k$ is the light wavelength, and $z_c\equiv \pi\rho_c^2/\lambda$ 
is the Rayleigh length (half-length of quasi-cylindrical region of the Gaussian beam). 
As was already noted, polarimetric signal in the single-beam arrangement (denote it $u_1$) 
can be calculated using Eq. (\ref{13}), by setting in it ${\bf A_0^t(R)}={\bf A_0(R)}$.  
With allowance for (\ref{8}) and (\ref{14}), we have

                    \begin{equation}
    u_1=\sin 2\phi \hskip1mm {8\pi k W\over c}\hskip1mm 
\int {d^3{\bf R}\hskip 1mm G_z({\bf R})\over \rho^2(z)}\hskip1mm \exp \bigg [-{x^2+y^2\over \rho ^2(z)}\bigg ]
    \label{15}
    \end{equation}

Since gyrotropy of the sample is connected with its magnetization, temporal fluctuations of the 
latter give rise to fluctuations of the gyrotropy:  $G_z({\bf R})\rightarrow G_z({\bf R},t)$.  
In a typical experiment on spin noise spectroscopy, one observes the noise power spectrum of the gyrotropy ${\cal N}(\nu)$, 
which is determined by Fourier transform of correlation function of the polarimetric signal 
${\cal N}(\nu)=\int dt\langle u_1(t)u_1(0)\rangle e^{\imath \nu t}$.  Using Eq. (\ref{15}), 
 we obtain, for the gyrotropy noise power spectrum, the expression

\begin{widetext}
  \begin{equation}
      {\cal N}(\nu)=\sin ^22\phi \hskip1mm \bigg [{8\pi k W\over c}\bigg ]^2\times 
      \label{16}
      \end{equation}
 $$
 \times \int dt \hskip1mm e^{\imath \nu t}\int {d^3{\bf R}\hskip1mm d^3{\bf R'}
       \over \rho^2(z)\rho^2(z')}\hskip1mm \exp \bigg [-\bigg ({x^2+y^2\over \rho ^2(z)} +  {x'^2+y'^2\over \rho ^2(z')}\bigg )\bigg ]\langle G_z({\bf R},t) \hskip0.5mm G_z({\bf R'},0)\rangle 
  $$
\end{widetext}

Correlation function of the gyrotropy $\langle G_z({\bf R},t) \hskip0.5mm G_z({\bf R'},0)\rangle$ 
entering this equation is calculated on the basis of one or another model of the sample under study. 
Most frequently, the gyrotropy is implied to be created by ensembles of gyrotropic particles 
(e.g., paramagnetic atoms) and is described by the expression

   \begin{equation}
     G_z({\bf R},t)=\sum_i g_i(t)\delta ({\bf R-r}_i(t)),
     \label{17}
     \end{equation}

where $g_i(t)\delta ({\bf R-r}_i(t))$ is the contribution of $i$-th particle to the total 
gyrotropy of the sample and ${\bf r}_i(t)$ is the coordinate of the $i$-th particle that may be time-dependent. 
The function $g_i(t)$ can be considered proportional to magnetic moment of the $i$-th particle, 
with the propotionalityu factor being, generally, dependent on the frequency $\omega$ of the light beam.

\subsection{ The model of resting paramagnetic particles.}

\begin{figure}
 \centering
 \includegraphics[width=\columnwidth,clip]{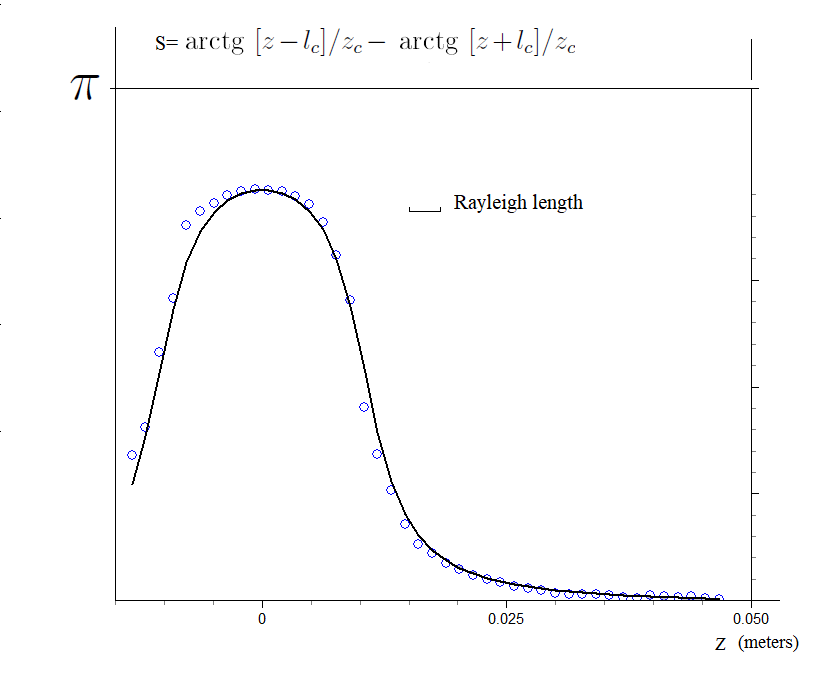}
 \caption{ Variation of area of the gyrotropy noise power spectrum ($S$) of Cs atoms in the Earth magnetic field with displacement of the cell ($z$) with respect to the light beam waist. 
 	 Solid curve -- theory, circles -- experiment.
 	  Wave length of the light beam and its waist radius are, respectively,  $\lambda=0.85$ nm  and $\rho_c=30 \mu$m.}
 \label{fig2}
\end{figure}

We start our treatment with the simplest model that implies that the sample consists of $N$ identical particles 
at rest, randomly distributed over the volume $V$ with the density $\sigma$ \cite{f3}. In this case, 
the gyrotropy is given by Eq. (\ref{17}) with time-independent coordinates of the particles ${\bf r}_i(t)\rightarrow {\bf r}_i$.
 The second assumption of this simple model is that the functions $g_i(t)$ are supposed to be random 
independent quantities, so that $\langle g_i(t)g_k(t')\rangle =\delta_{ik}\langle g(t-t')g(0)\rangle$.  
Here, the function $\langle g(t-t')g(0)\rangle$ is the same for all particles. 
Under these assumptions, for the correlator  entering Eq. (\ref{16}), we can obtain the following expression 
\cite{Koz}: $\langle G({\bf R},t)G({\bf R'},0)\rangle =\sigma \langle g(t)g(0)\rangle\hskip1mm \delta({\bf R-R'})$. 
 By substituting this expression into (\ref{16}) and calculating the integrals with $\delta$-functions, 
we obtain, for the noise power spectrum, the expression

 \begin{equation}
         {\cal N}(\nu)
                 =32\sigma \pi^3 \sin ^22\phi \hskip1mm \bigg [{   kW\over c}\bigg ]^2
                           \int dt \hskip1mm e^{\imath \nu t}\langle g(t) \hskip0.5mm g(0)\rangle \int {dz
                                  \over \rho^2(z)}
         \label{18}
         \end{equation}

Since  $1/\rho^2(z)\sim 1/[z^2+z_c^2]$ (see Eq. (\ref{14})), 
the main contribution to the signal is made by the region of the sample in the vicinity of the beam waist, where
$|z|<z_c$. This makes it possible to use the SNS method for tomography \cite{To},  
with the spatial resolution in the longitudinal direction, 
as expected, being determined by the Rayleigh length  $z_c$ of the probe beam.

Let us denote the bounds of the sample, along the light beam (i.e., along the $z$-axis) by $z_1$ and $z_2$. 
Then, using Eq. ({\ref{14}}) for the beam radius $\rho(z)$ and integrating over $z$ in Eq. ((\ref{18})), we obtain:

\begin{widetext}
\begin{equation}
           {\cal N}(\nu)=32 \sigma\pi^3 k^3 \sin ^22\phi \hskip1mm \bigg [{   W\over c}\bigg]^2
            \bigg [\hbox{ arctg }{z_2\over z_c}- \hbox{ arctg }{z_1\over z_c}\bigg ]
            \int dt \hskip1mm e^{\imath \nu t}\langle g(t) \hskip0.5mm g(0)\rangle 
           \label{19}
           \end{equation}
\end{widetext}

It follows from Eq. ((\ref{19})) that with increasing thickness of the sample 
($z_1\rightarrow -\infty$ and $z_2\rightarrow \infty$) the noise signal is saturated approaching the limiting value

 \begin{equation}
             {\cal N}_\infty (\nu)={32\pi^4\sigma k^3}\sin ^22\phi \hskip1mm \bigg [{   W\over c}\bigg]^2
              \int dt \hskip1mm e^{\imath \nu t}\langle g(t) \hskip0.5mm g(0)\rangle=
             \label{20}
             \end{equation}
$$
={32\pi^4\sigma k^3 T_2} \hskip1mm \bigg [{   W\over c}\bigg]^2
\bigg [{\sin ^2[2\phi]     
\langle g^2\rangle  \over 1+[\nu-\omega_L]^2 T_2^2 }+{\sin ^2[2\phi]     
\langle g^2\rangle \over 1+[\nu+\omega_L]^2 T_2^2 }\bigg ]
$$

 The last expression corresponds to  the correlator 
$\langle g(t) \hskip0.5mm g(0)\rangle=\langle g^2\rangle e^{-|t|/T_2}\cos\omega_Lt$ \cite{f5}.
To illustrate the above formulas, we have measured experimentally dependence of the noise signal area  
$\int {\cal N}(\nu) d\nu $ of cesium vapor on position $z$ of the cell  with respect to the 
beam waist (Fig. \ref{fig2}). 
The measurements were performed using a focused laser beam with the wavelength $\lambda =0.85    \hskip1mm \mu$m. 
The length of the cell $2l_c$  was 2 cm. In accordance with Eq. (\ref{19}), the measured dependence 
should have the form $\sim \hbox{ arctg }[z-l_c]/ z_c- \hbox{ arctg }[z +l_c]/ z_c $. 
As seen from Fig. \ref{fig2}, the experimental dependence is well approximated by 
this formula, with the best-fit value of the parameter $z_c$ ($z_c = 3.3 \cdot 10^{-3}$ m) 
well correlated with characteristics of the used laser beam. 
In spite of the fact that the cell thickness considerably exceeded the Rayleigh length $z_c$ 
(shown in Fig. \ref{fig2} by a horizontal segment), the value of the noise signal appeared 
to be noticeably (by $\sim 25 \%$) smaller than the limiting value (indicated in Fig. \ref{fig2} by a 
horizontal line at the level $\pi$). At the same time, it is seen from the presented experimental 
illustration and Eq. (\ref{19}) that, for the thickness of the sample $2l_c$ 
exceeding the Rayleigh length by a factor of 4-5, further reduction of the beam radius $\rho_c$ 
(with corresponding decrease of the Rayleigh length $z_c=\pi\rho_c^2/\lambda$) 
does not lead to substantial increase of the noise signal. Thus, it makes sense to decrease radius of the probe 
beam for increasing the value of the spin-noise signal only  for samples that are thin compared with the 
Rayleigh length of the light beam.

\subsection{The diffusion model}

Our assumption that the gyrotropy is created by resting particles is plausible for solid materials with  
embedded paramagnetic atoms giving rise to the gyrotropy.  For semiconductor samples, with the gyrotropy 
created by the moving charge carriers, as well as for gaseous systems, this assumption may be incorrect. 
It is natural to take into account the motion of gyrotropic particles in such systems using a diffusion 
model, with $N$ particles randomly moving in  a finite volume $V$ \cite{Prolet}. 
 In this case, Eq. (\ref{17}) for the gyrotropy 
remains valid.

Quantitative analysis and experimental    study of the diffusion effects    in SNS of gaseous systems has been recently presented in \cite{Luci}.    In this section, with the aid of relationships obtained above, we will reproduce the main results of \cite{Luci} treating the spin-noise signal as a result of scattering of a Gaussian probe beam. In addition, the notions introduced in this section will be used below to calculate the signal in the two-beam arrangement, when intuitive considerations about signal formation are not so self-evident as in conventional single-beam geometries.

If one considers a semiconductor system with a relatively low electron density in the 
conduction band or a gaseous system with diffusion motion of gyrotropic atoms occurring in a dense medium 
of nongyrotropic buffer gas, then contribution of each particle to gyrotropy of the sample can be considered 
as independent of other particles.  In this case, for the correlation function of gyrotropy, entering  
Eq. (\ref{16}) for the noise power spectrum, we can write the chain of equalities:
\begin{widetext}
  \begin{equation}
         \langle G_z({\bf R},t)G_z({\bf R}',0)\rangle=
         \sum_{ik}\langle g_i(t)g_k(0) \delta ({\bf R-r}_i(t))\delta ({\bf R'-r}_k(0))\rangle= 
         \label{21}
         \end{equation}
         $$
         =\langle g(t)g(0)\rangle\sum_{i}\langle \delta ({\bf R-r}_i(t))\delta ({\bf R'-r}_i(0))\rangle=N\langle g(t)g(0)\rangle\langle \delta ({\bf R-r}_1(t))\delta ({\bf R'-r}_1(0))\rangle=
         $$
         $$
         =N\langle g(t)g(0)\rangle\langle \delta ({\bf R-R'}-{\bf r}(t))\rangle \langle\delta ({\bf R'-r_1}(0))\rangle,
         $$
   \end{widetext}      
 
where  ${\bf r} (t)\equiv {\bf r_1}(t)-{\bf r_1}(0)$ -- is the vector of diffusion displacement of the particle 
 from the starting point ${\bf r}_1(0)$.  
 Here, we assume that fluctuations of gyrotropy for each particle are independent of its diffusion 
 motion \cite{f4} and suppose, as before, that 
 $\langle g_i(t)g_k(t')\rangle =\delta_{ik}\langle g(t-t')g(0)\rangle$. 
 Thus, the problem is reduced to studying diffusion motion of any single particle (e.g., the first one). 
The coordinate ${\bf r}_1(0)$ of this particle at $t = 0$ may acquire, with equal probability, 
any value within the volume $V$. Therefore, averaging of the last $\delta$-function over ${\bf r}_1$
 yields the factor $1/V$. In virtue of statistical uniformity of the sample, the distribution function $P({\bf r},t)$ 
of the vector of diffusion displacement ${\bf r} (t)\equiv {\bf r_1}(t)-{\bf r_1}(0)$ of the chosen particle 
does not depend on the starting point ${\bf r}_1(0)$ and is defined by the diffusion 
equation with the initial condition $P({\bf r},0)=\delta ({\bf r})$:

  \begin{equation}
   {\partial P\over\partial t}=D\Delta P  \hskip10mm P({\bf r}, 0)=\delta ({\bf r})
   \label{22}
   \end{equation}
   
where $D$ is the diffusion coefficient, 
${\bf r}=(x,y,z)$ and $\Delta=\partial^2/\partial x^2+\partial^2/\partial y^2+\partial^2/\partial z^2$ -- 
is the Laplace operator. Thus, the chain of equalities (\ref{21}) can be continued as follows:

   \begin{equation}
         \langle G({\bf R},t)G({\bf R}',0)\rangle= 
         {N\over V}\langle g(t)g(0)\rangle\langle \delta ({\bf R-R'}-{\bf r}(t))\rangle =         
         \label{23}
         \end{equation}
   $$
  = \sigma\langle g(t)g(0)\rangle\langle \delta ({\bf R-R'}-{\bf r}(t))\rangle =\sigma\langle g(t)g(0)\rangle P({\bf R-R'},t)
   $$
 
 here  $\sigma=N/V$ is the density of the particles.
Standard solution of the problem (\ref{22}) leads to the following expression 
 for the distribution function $P({\bf r},t)$:
   \begin{equation}
   P({\bf r},t)={1\over 8(\pi Dt)^{3/2}}\exp\bigg [-{r^2\over 4Dt} \bigg ],
   \label{24}
   \end{equation}
 
Substituting this function into (\ref{23}), we obtain, for the gyrotropy correlator in the presence 
of diffusion, the final expression:

   \begin{equation}
    \langle G_z({\bf R},t)G_z({\bf R}',0)\rangle={\sigma\langle g(t)g(0)\rangle
    \over 8(\pi D|t|)^{3/2}}\exp\bigg [-{|{\bf R-R'}|^2\over 4D|t|} \bigg ] 
         \label {25}
         \end{equation}
        
  Here, we took into account parity of the correlation function. By substituting this expression into Eq. (\ref{16}) for the noise power spectrum we obtain

       \begin{equation}
             {\cal N}(\nu)=
             \sin ^22\phi \hskip1mm {8\sqrt{\pi}\sigma\over D^{3/2}} \bigg [{ k W\over c}\bigg ]^2\int dt \hskip1mm e^{\imath \nu t}{\langle g(t)g(0)\rangle
                  \over  |t|^{3/2}}\times 
             \label{26}
             \end{equation}
     $$
    \times\int {d^3{\bf R}\hskip1mm d^3{\bf R'}
          \over \rho^2(z)\rho^2(z')}\hskip1mm \exp \bigg [-{x^2+y^2\over \rho ^2(z)} -  {x'^2+y'^2\over \rho ^2(z')}\bigg ]\exp\bigg [-{|{\bf R-R'}|^2\over 4D|t|} \bigg ]
          $$
   Here, ${\bf R}=(x,y,z)$ and ${\bf R'}=(x',y',z')$.  The integrals over $x,y,x',y'$ are reduced to Gaussian by appropriate rotations of the coordinate system in the planes $xy$ and $x’y'$,  that eliminate in the exponent the terms $\sim xy$ and $\sim x’y’$. By calculating these Gaussian integrals, we come to the following expression for the noise power spectrum:
      \begin{widetext}
       \begin{equation}
       {\cal N}(\nu)
       ={32\sigma \pi^{5/2}\over  D^{1/2}}\hskip1mm\sin ^22\phi \hskip1mm \bigg [{ k W\over  c}\bigg ]^2\int {dt\over |t|^{1/2}} \hskip1mm {e^{\imath \nu t}
         }
         \int {dz dz' \langle g(t)g(0)\rangle\over 4D|t|+\rho^2(z)+\rho ^2(z')}
                \exp\bigg [-{( z-z')^2\over 4D|t|} \bigg ],
       \label{27}
        \end{equation}
        \end{widetext}
 that transforms  to Eq. (\ref{18}) at $D\rightarrow 0$.

Equation (\ref{27}) can be simplified assuming that the diffusion length for the characteristic decay time of the correlator  $\langle g(t)g(0)\rangle $ is smaller than the Rayleigh length. In the situation typical for the SNS, when the correlator $\langle g(t)g(0)\rangle $ decreases exponentially, $\langle g(t)g(0)\rangle=\langle g^2\rangle e^{-|t|/T_2}\cos \omega_Lt$ \cite{f5}, the above condition can be written in the form: $\sqrt {DT_2}\ll z_c$ (see Eq. (\ref {14})). In this case, we may put, in Eq. (\ref{27}), $\rho(z)\approx\rho(z')$, perform integration over $z'$ , and obtain the following simplified expression for the noise power spectrum:

  \begin{equation}
  {\cal N}(\nu)
  =32\pi^3\sigma\hskip1mm\sin ^22\phi \hskip1mm \bigg [{ k W\over c}\bigg ]^2  \int dtdz \hskip1mm {e^{\imath \nu t}\langle g(t)g(0)\rangle
    \over 2D|t|+\rho^2(z)} 
  \label{28}
   \end{equation}
  $$
  \hskip5mm \hbox{ at } \hskip2mm 2\sqrt{DT_2}<z_c,
  $$
   
   where $\rho(z)$ is defined by Eq. (\ref{14}). It is seen from this relationship that, in the region of the sample where $\rho(z)<\sqrt{2DT_2}$ (provided that such a region exists), time dependence of the integrand deviates from $\sim\langle g(t)g(0)\rangle $ that is usually exponential. As a result, the shape of the noise power spectrum is deviated from Lorentzian, and  the noise spectrum reveals  the so-called time-of-flight broadening \cite{Prolet}. If the beam is so broad that $\rho_c>\sqrt{DT_2}$, then this effect proves to be suppressed and  can be neglected. Estimates show that conditions of applicability of Eq. (\ref{28}) often come true in practice.  Using Eq. (\ref{14}) for the function $\rho(z)$, the integration over $z$ in Eq. (\ref{28}) can be performed analytically.  Let us present the result for the case when the sample length is much larger than both the Rayleigh length and the  diffusion length  $\sqrt {DT_2}$  for the time $T_2$:
   
   \begin{equation}
    {\cal N}(\nu)=32 \pi ^4k^3\rho_c\hskip1mm \sigma \sin ^2 2\phi \bigg ({W\over c}\bigg )^2\int dt {e^{\imath \nu t}\langle g(t)g(0)\rangle\over \sqrt{4D|t|+\rho_c^2}}
    \label{29}
    \end{equation}
    $$
    \hskip5mm \hbox{ at } \hskip2mm   2\sqrt{DT_2}\ll z_c 
        \hskip2mm \hbox{ and } l_s\gg z_c
        $$
  As seen from Eq. (\ref{29}), when the diffusion drift $\sqrt{DT_2}$ for the time $T_2$ is smaller  than the beam radius $\rho_c$, the effects of diffusion can be neglected. Otherwise, the noise spectrum exhibits the time-of-flight broadening.

 \section{The two-beam noise spectroscopy}

 Above, we presented calculations of the noise signals detected in the single-beam arrangement, traditional for the SNS. Consider now the case when the beam that induces scattering and the beam that plays the role of local oscillator are different 
 ${\bf A_0(R)\ne A_0^t(R)}$ \cite{Koz} (Fig.\ref{fig1}).  We will assume that waists of these two beams intersect in the region of the studied gyrotropic sample and will analyze the problem under the following simplifying assumptions:

 (i)	Both beams propagate in the direction close to the $z$-axis, the angle $\Theta$ between the beams is {\it small enough} to make possible low-power approximations of its trigonometric functions, and main components of electric fields of the beams lie in the $xy$ plane.

 (ii)	The angle $\Theta$ is {\it large enough} not to make length of the beam overlap larger than the Rayleigh length.

 Appropriate quantitative conditions will be presented below. Let us choose the coordinate system so that both the beams (the main and auxiliary) lie in the plane $yz$. (i.e., the beams are rotated with respect to each other around the $x$-axis). Bearing in mind the first of the above assumptions, polarizations of the main and auxiliary beams are specified by the following two-dimensional (in the plane $xy$) Jones vectors:
 
 \begin{equation}
 {\bf d}=\left (\begin{matrix} \cos\phi \cr \sin\phi \end{matrix}\right )\hskip10mm 
 {\bf d^t}=\left (\begin{matrix}\cos\eta \cr \sin\eta \end{matrix}\right )
 \end{equation}

 Using the second of the above assumptions, we can neglect, in Eqs. (\ref{8}) and (\ref{9}), the terms $Q^2z$ and $Q^2Z$ as compared with $2k$.  After that, with the aid of Eq. (\ref{13}), we obtain, for the polarimetric signal produced by the auxiliary beam (below referred to as  $u_1^t(t)$),  the following expression ($\delta U_t \rightarrow u_1^t(t)$):

  \begin{equation}
     u_1^t(t)= {16\pi k \over\rho_c^2 c}\sqrt {WW_t}\int d^3{\bf R}\cos \bigg (k\Theta y+{k\Theta^2z\over 2}-\phi_t\bigg )\times 
     \label{31}
     \end{equation}
  $$
  \times \sin [\phi+\eta] \exp \bigg [-2\hskip1mm {   
          x^2+y^2+yz\Theta+z^2\Theta^2/2\over\rho_c^2 }\bigg ]\hskip1mm G_z({\bf R},t),
  $$

  When deriving this formula, we took into account smallness of the angle $\Theta$ (see transformations (\ref{10})). Exponential of the quadratic form of the coordinates, in this formula, is essentially nonzero in the region  $\sim \rho_c\times\rho_c$   (in the plane $xy$) over the length $\sim \rho_c/\Theta$  (along the $z$-axis) . Therefore, the second of the above assumptions can be expressed by the inequality $\rho_c/\Theta<\pi\rho_c^2/\lambda$. Keeping it in mind, we come to conclusion that the above assumptions impose the following restrictions upon the angle $\Theta $ between the beams:
   \begin{equation}
   {\lambda\over \pi\rho_c}<\Theta <1
   \label{266}
   \end{equation}

 Typically, $\rho_c\sim 30 \hskip1mm \mu$m  at $\lambda\approx 1 \mu$m.  Therefore, for validity of the calculations carried out in this Section,  the angle $\Theta$ should meet the inequality $ 10^{-2}<\Theta<1$, that can be easily satisfied in practice.  
 
 When calculating the noise power spectrum detected in the two-beam arrangement, one has to take into account that the total polarimetric signal $\delta U(t)$, in this case, is the sum: $\delta U(t)=u_1(t)+u_1^t(t)$,  $u_1(t)$ and $u_1^t(t)$ given by Eqs. ((\ref{15}) and (\ref{31})), respectively. Hence, the formula for the noise power spectrum  ${\cal N}(\nu)=\int   e^{\imath \nu t}\langle \delta U(0)\delta U(t)\rangle dt$ will contain four contributions:

 \begin{equation}
    {\cal N}(\nu)=\int dt \bigg [\langle  u_1(t) u_1(0)\rangle+\langle  u_1(t) u_1^t(0)\rangle+
    \label{33}
    \end{equation}
 $$
 +\langle  u_1^t(t) u_1(0)\rangle+\langle  u_1^t(t) u_1^t(0)\rangle \bigg ] e^{\imath \nu t} 
 $$

   If no special measures are taken to stabilize relative phase $\phi_t$ of the main and auxiliary beams, then it is natural to perform averaging over this phase, which will be below implied. As a result of this averaging, the cross-correlators $\langle u_1^t(0)u_1(t)\rangle$ and $\langle  u_1^t(t) u_1(0)\rangle$  will vanish. The first correlator $\langle  u_1(t) u_1(0)\rangle $  in Eq. (\ref{33})  has been already calculated  above  (Eq. (\ref{16})). It gives the noise spectrum observed in the single-beam arrangement. For this reason, in what follows we will consider the contribution to the noise spectrum related only to the auxiliary beam and controlled by the correlator $\langle  u_1^t(t) u_1^t(0)\rangle $. Let us denote this contribution by ${\cal N}_t(\nu)\equiv \int dt \langle  u_1^t(t) u_1^t(0)\rangle  e^{\imath \nu t}$. If the sample gyrotropy represents a random field statistically stationary in space and in time, then its correlation function depends only on difference between its spatiotemporal arguments and can be represented in the form ${\cal K}({\bf R-R'},t)\equiv \langle G({\bf R},t)G({\bf R}',0)\rangle$.  Using Eq. (\ref{31}), we obtain for the correlator $\langle  u_1^t(t) u_1^t(0)\rangle$ the following relation:
   \begin{widetext}
    \begin{equation}
      \langle  u_1^t(t) u_1^t(0)\rangle={128\pi^2 k^2 \over\rho_c^4 c^2}WW_t
      \sin ^2[\phi+\eta]\int d^3{\bf R} d^3{\bf R}'
      \exp \bigg [-2\hskip1mm {   
               x^2+y^2+yz\Theta+z^2\Theta^2/2\over\rho_c^2 }\bigg ]\times
           \label{34}
      \end{equation}
     $$
      \times\exp \bigg [-2\hskip1mm {   
                x'^2+y'^2+y'z'\Theta+z'^2\Theta^2/2\over\rho_c^2 }\bigg ]
     \cos \bigg [k\Theta (y-y')+{k\Theta^2(z-z')\over 2}\bigg ] {\cal K}({\bf R-R'},t)
     $$           
     \end{widetext}
 Here, the averaging over the relative phase of the beams $\phi_t$ is performed. 
 
 Using Eq.(\ref{34}) as a starting point, we can obtain a simpler approximate formula, suitable for estimating the SNS signals under experimental conditions typical for this method.  Note that the exponential factors in Eq. (\ref{34}), in fact, shrink the integration region to the region of overlap between the main and auxiliary beams.  The volume of this region $V_o$ can be evaluated in the following way:

          \begin{equation}
            V_o\approx \int d^3{\bf R}\exp \bigg [-2\hskip1mm {   
                        x^2+y^2+yz\Theta+z^2\Theta^2/2\over\rho_c^2 }\bigg ]=
            \label{35}
            \end{equation}
        $$
        ={\pi^{3/2}\rho_c^3\over \sqrt 2\Theta}.
        $$
  
  For this reason, in Eq. (\ref{34}), we may restrict the region of integration over 
   $d^3{\bf R}$ and $d^3{\bf R'}$
   with the volume $V_o$ and set the exponential factors to be equal to unity.  After that, the integrand will appear to be dependent on the difference ${\bf R-R'}$. Now, let us pass to new variables ${\bf r\equiv R-R'}$ and ${\bf g\equiv R+R'}$.  The integral over ${\bf g}$ will give the volume of integration $V_o$, and for the correlator (\ref{34}) we can write the following approximate formula 
  \begin{widetext}
    \begin{equation}
     \langle  u_1^t(t) u_1^t(0)\rangle\approx  \ae {\pi^{7/2} k^2 \over\rho_c c^2}WW_t
        \sin ^2[\phi+\eta]    \int_{V_o} d^3{\bf r} 
       \cos  (\Delta {\bf k, r} ) {\cal K}({\bf r},t),\hbox{ where }\hskip5mm \Delta {\bf k}\equiv k\Theta \left (\begin{matrix} 0\cr 1\cr \Theta /2 \end{matrix}\right )
             \label{36}
    \end{equation}
    \end{widetext}
     is the difference wave vector of the main and tilted beams, while the numerical factor is $\ae={16/ \sqrt 2}$. When the region of overlap of the beams is large compared with the gyrotropy correlation radius and spatial periods of cosine 
   in (\ref{34}) -- $\lambda/2\pi\Theta$
     (in the $y$-direction) and $\lambda/2\pi\Theta^2$ (in the $z$-direction), then the integral in (\ref{36}) coincides with the Fourier transform of the gyrotropy correlation function. 
   
   Using  relation (\ref{36}), we can calculate the contribution to the gyrotropy noise power spectrum associated with the auxiliary beam in the presence of diffusion.  For the correlation function of the gyrotropy, we use Eq. (\ref{25}), in which we set 
   $\langle g(t)g(0)\rangle =\langle g^2\rangle e^{-|t|/T_2}\cos \omega_Lt$ \cite {f5}. Calculating the Fourier transform of Eq. (\ref{25})   and substituting it to Eq. (\ref{36}), we obtain

   \begin{widetext}   
    \begin{equation}
                 {\cal N}_t(\nu)\approx \ae  \hskip1mm { T_2^\ast\sigma\pi^{7/2} k^2 \over\Theta\rho_c c^2}WW_t
                 \hskip2mm 
                    \bigg [{\sin ^2[\phi+\eta]     
\langle g^2\rangle  \over 1+[\nu-\omega_L]^2 T_2^{\ast 2} }+{\sin ^2[\phi+\eta]     
\langle g^2\rangle \over 1+[\nu+\omega_L]^2 T_2^{\ast 2} }\bigg ],
                         \label{37}
                \end{equation}
                \end{widetext}
        where 
        \begin{equation}
        T_2^\ast \equiv {T_2\over 1+k^2\Theta^2 DT_2} 
        \label{38}
        \end{equation}

As seen from this formula,   diffusion leads to broadening of the noise spectrum and reduction of its amplitude, 
provided that the diffusion length for the dephasing time $\sqrt{DT_2}$ exceeds spatial period of 
 interference between the main 
and auxiliary beams $1/k\Theta =\lambda/ 2\pi\Theta$.  In the opposite case 
(i.e., at $\sqrt{DT_2}< \lambda/ 2\pi\Theta$), the contributions ${\cal N}_t(\nu)$ (\ref{37})  
and ${\cal N}(\nu)$ (\ref{27}) have comparable 
amplitudes and spectral widths. 

Now, let us present arguments that allow us to believe that  Eq. (\ref{37}) works well even when the basic conditions 
of its derivation are satisfied poorly.  For this, we present the result of consistent computation of integral (\ref{34})
 with the correlation function of gyrotropy in the form (\ref{25}). In this case, the integrand represents an exponential 
of some quadratic forms of the integration variables. Such a form can be diagonalized with the proper orthogonal 
transformation of coordinate system. After this, integral (\ref{34}) is reduced to a product of Gaussian integrals.  
Omitting cumbersome manipulations, we present final result of such calculations:

\begin{widetext}
   \begin{equation}
           \langle  u_1^t(t) u_1^t(0)\rangle=
          {8\pi^{7/2} k^2 \over\rho_c^2 c^2}WW_t
             \sin ^2[\phi+\eta]
             {\sigma\langle g(t)g(0)\rangle
               \over ( Dt)^{3/2}}
              \bigg [1+{\rho_c^2\over 4Dt}\bigg ]^{-1/2}
              {\exp [-(M^{-1}h,h)/4]\over \sqrt{\hbox {det }M} }, 
           \label{39}
             \end{equation}  
            \end{widetext}   
where the vector-column $h$ and the matrix $M$ are defined by the relations:

    \begin{equation}
    h=k\Theta \left (\begin{matrix}1\cr \Theta/2\cr -1\cr -\Theta/2\end{matrix}\right )
    \end{equation}
    \begin{widetext}
     \begin{equation}
      M\equiv\left (\begin{matrix} \alpha & \delta & \gamma & 0 \cr
      \delta &\beta & 0 &\gamma \cr
      \gamma & 0 & \alpha & \delta \cr
      0 & \gamma & \delta & \beta \end{matrix}\right )\hskip 5mm
      \alpha \equiv {2\over \rho_c^2}+{1\over 4Dt}, 
      \hskip5mm \beta \equiv {\Theta ^2\over \rho_c^2}+{1\over 4Dt}, \hskip5mm
      \delta \equiv {\Theta \over \rho_c ^2}, \hskip5mm \gamma\equiv -{1\over 4Dt}
      \label{41}
      \end{equation}
\end{widetext}
Calculations of the correlation functions of the polarimetric signal show that the results obtained using 
(\ref {37}) and  (\ref {39}) at $\rho_c>3\lambda$ 
and $0.05<\Theta<0.3$ 
 practically coincide if we set in Eq. (\ref {37})  $\ae=32$. 

Our efforts to observe the noise signal from cesium atoms (see the end of Sect. 2.1) associated with the 
auxiliary beam (Fig. \ref{fig1}) have failed. The reason of this failure is likely to be the following. 
Let us compare amplitude of the noise signal (\ref{37}) related to the auxiliary beam with that of signal 
(\ref{20}) detected in the single-beam arrangement. Using Eqs. (\ref{37}) and (\ref{20}), at $\omega_LT_2\gg 1$,
  we obtain the relationship

\begin{equation}
{{\cal N}_t(\omega_L)\over {\cal N}_\infty (\omega_L)}={1\over 1+ k^2\Theta^2DT_2}\hskip2mm
{\lambda\over 2\pi^{3/2}\rho_c\Theta}\hskip2mm{W_t\over W}\hskip2mm{\sin^2[\phi+\eta]\over \sin^ 22\phi}
\end{equation}

The two last factors can be made $\sim 1$ by tuning polarization and intensities of the main and auxiliary beams. 
The second factor describes decrease of the noise signal in the two-beam arrangement resulted from 
incomplete overlap of the two beams. At 
$\lambda\sim 1\hskip1mm \mu$m, $\rho_c\sim 30\hskip1mm  \mu$m  and $\Theta\sim 0.1$ rad, this factor is $\sim$ 1/30. And, 
finally, the first factor describes decrease of the noise spectrum amplitude ${\cal N}_t(\nu)$ 
associated with diffusion of the gyrotropic particles. Let us estimate this factor for our particular 
case of cesium atoms. 
Taking for the diffusion coefficient of Cs atoms in the buffer gas atmosphere the 
value  $D= 2\cdot 10^{-5}$ m$^2/$sec \cite{Diff}  and for the dephasing time of Cs spins the value 
$T_2\sim 10^{-3}-10 ^{-4}$ sec \cite{f6},  we obtain that, at 
$\Theta=0.1$ and $\lambda=1 \mu$m, the quantity $k^2\Theta ^2DT_2$ is $\sim 10^3$.  
Thus, in our case, the noise signal associated with the tilted beam appears to 
be suppressed by a factor of $\sim 3\cdot 10^4$  that substantially hampers its detection. 
It seems  that observation of this signal may appear possible for systems with weak diffusion 
(like quantum dots) or for semiconductor systems with shorter dephasing time $T_2$, 
when the noise signal from quasi-free electrons, in the single-beam arrangement, can still be reliably detected.

 \section*{Conclusions}

In this paper, we perform, in the single-scattering approximation, consistent calculations of polarimetric 
signal detected in the spin noise spectroscopy (SNS). The derived expressions can be applied to samples 
with the length exceeding that of Rayleigh of the probe laser beams.  The calculations are performed for 
model systems comprised of gyrotropic particles with allowance for their possible diffusion. 
Analysis of two-beam arrangement of the SNS is presented that makes it possible to study not only temporal, 
but also spatial correlations of the gyrotropy. It is shown that diffusion of gyrotropic particles may 
broaden the noise spectra observed in the two-beam arrangement, with this broadening substantially exceeding 
the time-of-flight  broadening observed in conventional single-beam arrangement.

\section*{Acknowledgements}
The authors appreciate support of Russian Science Foundation (Project no.17-12-01124).
  The work was carried
out using the equipment of SPbU Resource Center
"Nanophotonics" (photon.spbu.ru).


\begin{thebibliography}{99}
\bibitem{f1} This  is performed by measuring polarisation modulation of the probe beam passing through the sample under study
\bibitem{f0} Its noteworthy that the function  $\Phi({\bf r})$ has the sence of the field created by the source that is distributed over the surface of the detector and has the "density"  ${\bf {\cal E}}_0(x,y,L)$. For this reason,   the field $\Phi({\bf r})$ similar to that of the main beam
  ${\bf {\cal E}}_0({\bf r})$. Explicit expression for $\Phi({\bf r})$ was derived  \cite{Koz1}
   and is presented below Eq. (\ref{12}).
\bibitem{f3} Such a model, for samples thin compared with the Rayleigh length, was considered in \cite{Koz}.
\bibitem{f5}   Such a correlator corresponds to the case when gyrotropy of the sample is created by the system of paramagnetic particles in a magnetic field; the quantities $\omega_L$  and $T_2$ are, respectively, the Larmor frequency of the effective spin and its dephasing time.
\bibitem{f4} 
 This assumption is arguable but for the simplest analysis of the diffusion effects is, in our opinion, acceptable.
\bibitem{f6}  This value is obtained from our observations of the noise spectrum of Cs atoms.
\end{thebibliography}
\end{document}